\documentclass[letterpaper, 10pt, conference]{ieeeconf}
\IEEEoverridecommandlockouts 
\usepackage{amssymb}
\usepackage{amsmath}

\usepackage{amsthm}
\usepackage{cite}
\usepackage{graphics}
\usepackage{hyperref}
\usepackage{multirow}
\usepackage{tikz}

\usetikzlibrary{shapes,arrows}
\tikzstyle{block} = [draw, rectangle]
\tikzstyle{sum}   = [draw, circle]
\tikzstyle{point} = [coordinate]

\theoremstyle{definition}
\newtheorem{defn}{Definition}
\newtheorem{theorem}{Theorem}

\def\R{\mathbf{\Phi}_x}
\def\M{\mathbf{\Phi}_u}
\def\bx{\mathbf{x}}
\def\bu{\mathbf{u}}
\def\bw{\mathbf{w}}
\def\bK{\mathbf{K}}
\def\Ahat{\hat{A}}
\def\Bhat{\hat{B}}

\newif\ifshowWriterComment
\newcommand\writercomment[3]{\expandafter\newcommand\csname #2\endcsname[1]{\ifshowWriterComment{\color{#3} #1: ##1}\fi}}

\writercomment{Lisa}{Lisa}{purple}
\writercomment{Carmen}{Carmen}{blue}

\showWriterCommenttrue

\title{\LARGE \bf
    Frontiers in Scalable Distributed Control: SLS, MPC, and Beyond
}

\author{Jing Shuang (Lisa) Li, Carmen Amo Alonso, and John C. Doyle
	\thanks{Authors are with the Department of Computing and Mathematical Sciences, California Institute of Technology.
		{\tt\small \{jsli, camoalon, doyle\}@caltech.edu},
	}%
}
{\tiny }
\begin{document}	
	\maketitle
	\thispagestyle{empty}
	\pagestyle{empty}
	
	\begin{abstract}
	
The System Level Synthesis (SLS) approach facilitates distributed control of large cyberphysical networks in an easy-to-understand, computationally scalable way. We present an overview of the SLS approach and its associated extensions in nonlinear control, MPC, adaptive control, and learning for control. To illustrate the effectiveness of SLS-based methods, we present a case study motivated by the power grid, with communication constraints, actuator saturation, disturbances, and changing setpoints. This simple but challenging case study necessitates the use of model predictive control (MPC); however, standard MPC techniques often scales poorly to large systems and incurs heavy computational burden. To address this challenge, we combine two SLS-based controllers to form a layered MPC-like controller. Our controller has constant computational complexity with respect to the system size, gives a 20-fold reduction in online computation requirements, and still achieves performance that is within 3\% of the \textit{centralized} MPC controller.

\end{abstract}
	\section{INTRODUCTION} \label{sec:introduction}

The control of large cyberphysical systems is important to today's society. Relevant examples include power grids, traffic networks, and process plants. In each of these fields, emerging technology (e.g. renewables, autonomous vehicles) and increasingly complex architectures present new challenges, and theoretical frameworks and algorithms are needed to address them. Generally speaking, these widespread large-scale systems require a distributed control framework that offers scalable, structured solutions.

The recently introduced System Level Synthesis (SLS) framework \cite{Anderson2019} provides theory and algorithms to deal with large, complex, structured systems. SLS tackles common challenges in distributed control, including disturbance containment and communication constraints. Moreover, it enables distributed synthesis \textit{and} implementation; the runtime of the synthesis algorithm is independent of the network size, and each subsystem can synthesize its own controller, bypassing the need for centrally coordinated controller synthesis and making this framework highly scalable.

Since its inception, many extensions of the SLS framework have been developed, including works on nonlinear plants \cite{Ho2020_nonlinear, Yu2020_saturation}, model predictive control (MPC) \cite{AmoAlonso2019_mpc, AmoAlonso2020_explicitmpc}, adaptive control \cite{Ho2019_adaptive, Han2020_learning}, and learning \cite{Matni2019_rl, Dean2019_safety, Dean2020_perception}; the core SLS ideas are useful and applicable to a variety of settings. In this paper, we hope to familiarize more researchers and practitioners with this powerful and scalable framework. 

One notable extension of the SLS framework is scalable MPC. MPC is ubiquitous in industry, but challenging for use in large networks due to scalability issues and high computational demand. Various distributed and hierarchical methods have been proposed to address this \cite{Scattolini2009}; in this paper, we propose a novel SLS and MPC-based layered controller that is high-performing and uniquely scalable. We demonstrate our controller on an example system motivated by a power grid, which is subject to communication constraints, actuator saturation, disturbances, and setpoint changes that result from intermittently shifting optimal power flows (OPFs).

This paper introduces the basic mathematics of SLS (Section \ref{sec:sls_math}), then presents a comprehensive overview of all SLS-based methods to date (Section \ref{sec:sls_techniques}). These sections form a useful introductory reference for  the system-level approach to controls. We follow up with a SLS-based case study (Section \ref{sec:case_study}) in which we introduce a scalable distributed two-layered controller that successfully approximates centralized MPC performance while greatly reducing online computation.

	\section{THE SLS PARAMETRIZATION} \label{sec:sls_math}

We introduce the basic mathematics of SLS, adapted from $\S2$ of \cite{Anderson2019}. For simplicity, we focus on the finite-horizon state feedback case; analogous results for infinite-horizon and output feedback can be found in $\S4$ and $\S5$ of \cite{Anderson2019}. We also briefly present the SLS-based MPC formulation.

\subsection{Setup}
Consider the discrete-time linear time varying (LTV) plant
\begin{equation} \label{eqn:LTV_system}
x(t+1) = A_tx(t)+B_tu(t)+w(t),
\end{equation}

\noindent where $x(t)\in\mathbb{R}^{n}$ is the state, $w(t)\in\mathbb{R}^{n}$ is an exogenous disturbance, and $u(t)\in\mathbb{R}^{p}$ is the control input. The control input is generated by a causal LTV state-feedback controller

\begin{equation} \label{eqn:LTV_ctrller}
u(t) = K_t(x(0), x(1), ..., x(t))
\end{equation}

\noindent where $K_t$ is some linear map. Let $Z$ be the block-downshift operator\footnote{Sparse matrix composed of identity matrices along its first block sub-diagonal.}. By defining the block matrices $\Ahat :=\mathrm{blkdiag}(A_1,A_2,...,A_T,0)$ and $\Bhat :=\mathrm{blkdiag}(B_1,B_2,...,B_T,0)$, the dynamics of system \eqref{eqn:LTV_system} over the time horizon $t = 0, 1, ..., T$ can be written as

\begin{equation} \label{eqn:LTV_freq}
	\bx = Z \Ahat \bx + Z \Bhat \bu + \bw
\end{equation}

\noindent where $\bx$, $\bu$, and $\bw$ are the finite horizon signals corresponding to state, disturbance, and control input respectively. The controller \eqref{eqn:LTV_ctrller} can be written as 

\begin{equation} \label{eqn:LTV_controller}
	\bu = \bK \bx
\end{equation}

\noindent where $\bK$ is the block-lower-triangular (BLT) matrix corresponding to the causal linear map $K_t$.

\subsection{System Responses}
Consider the closed-loop system responses $\left\{\R, \M\right\}$, which map disturbance to state and control input, i.e.
\begin{subequations} \label{eqn:closedloopmaps}	
	\begin{equation} 
	\mathbf{x} := \R \mathbf w
	\end{equation}
	\begin{equation}
	\mathbf{u} := \M \mathbf w
	\end{equation}
\end{subequations}

By combining \eqref{eqn:LTV_freq} and \eqref{eqn:LTV_controller}, we easily see that

\begin{subequations} \label{eqn:Phis}
	\begin{equation}
	\R = (I-Z(\Ahat + \Bhat \bK))^{-1}
	\end{equation}
	\begin{equation}
	\M = \bK (I-Z(\Ahat + \Bhat \bK))^{-1} = \bK \R
	\end{equation}
\end{subequations}

\begin{defn}
	$\left\{\R, \M\right\}$ are \textit{achievable} system responses if $\exists$ a block-lower-triangular matrix $\bK$ (i.e. causal linear controller) such that $\R$, $\M$, and $\bK$ satisfy \eqref{eqn:Phis}. If such a $\bK$ exists, we say that it \textit{achieves} system responses $\left\{\R, \M\right\}$.
\end{defn}

The SLS framework works with system responses $\left\{\R, \M\right\}$ directly. We use convex optimization to search over the set of achievable $\left\{\R, \M\right\}$, then implement the corresponding $\bK$ using $\left\{\R, \M\right\}$. We can do this because the set of achievable system responses is fully parametrized by an affine subspace, as per the core SLS theorem:

\begin{theorem}{\label{thm:SLS}}
For plant (\ref{eqn:LTV_system}) using the state-feedback policy $\bu = \bK \bx$, where $\bK$ is a BLT matrix, the following are true
\begin{enumerate}
    \item The affine subspace of BLT $\left\{\R, \M\right\}$
    \begin{equation}\label{eqn:SLS_constraint}
    	\begin{bmatrix}
    		I - Z\Ahat && -Z\Bhat
    	\end{bmatrix}
    	\begin{bmatrix}
    		\R \\ \M
    	\end{bmatrix}
    	= I
    \end{equation}
    parametrizes all achievable system responses \eqref{eqn:Phis}.
    
    \item For any BLT matrices $\left\{\R, \M\right\}$ satisfying \eqref{eqn:SLS_constraint}), the controller $\bK = \M\R^{-1}$ achieves the desired response \eqref{eqn:Phis} from $\mathbf w \mapsto (\mathbf x,\mathbf u)$.
\end{enumerate}
\end{theorem}

{\begin{proof}See Theorem 2.1 of \cite{Anderson2019}. \end{proof}}

Theorem \ref{thm:SLS} allows us to reformulate an optimal control problem over state and input pairs $(\bx, \bu)$ as an equivalent problem over system responses $\left\{\R, \M\right\}$. As long as the system responses satisfy \eqref{eqn:SLS_constraint} (which can be interpreted as a generalization of controllability), part 2 of Theorem \ref{thm:SLS} guarantees that we will also have a controller $\bK$ to achieve these system responses. Thus, a general optimal control problem can be formulated in the SLS framework as
\begin{equation} \label{eqn:synthesis}
\begin{array}{rl}
\underset{\R, \M}{\text{min}} & f(\R, \M) \\
 \text{s.t.} &  \eqref{eqn:SLS_constraint}, \, \R \in \mathcal{X}, \M \in \mathcal{U},
\end{array}
\end{equation}
where $f$ is any convex objective function and $\mathcal{X}$ and $\mathcal{U}$ are convex sets. Details of how to choose $f$ for several standard control problems is provided in $\S2$ of \cite{Anderson2019}. Common specifications for distributed control, such as disturbance containment, communication delay, local communication, and actuation delay, can be enforced by sparsity patterns on $\mathcal{X}$ and $\mathcal{U}$. Suitable specifications are discussed at length in \cite{Anderson2019}.

For many classes of $f$ (e.g. $\mathcal{H}_2$ objective), \eqref{eqn:synthesis} decomposes into smaller subproblems that can be solved in parallel. This allows for \textit{distributed synthesis}; each subsystem in the system solves a local subproblem independently of other subsystems. Under localized communication constraints, the size of these subproblems are independent of total system size $N$, making the SLS formulation uniquely scalable. The size of these subproblems depend on $d$, the size of the local region; $d$-localized communication constrains each subsystem to only communicate with subsystems at most $d$-hops away. For large distributed systems, generally $d \ll N$, so scaling with $d$ instead of $N$ is highly beneficial.

\subsection{Controller Implementation}
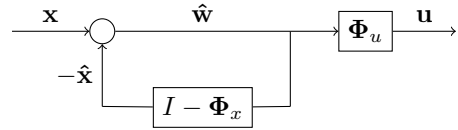
\begin{figure} 
	\centering
	\begin{tikzpicture}
	\node [point, name=input] {};
	\node [sum, right of=input, node distance=1.2cm] (sum) {};
	\node [point, right of=sum, node distance=2.5cm] (pt1) {};
	\node [block, right of=sum, node distance=3.5cm] (zM) {$\M$};
	
	\node [point, right of=zM, node distance=1.2cm, name=output] {};
	
	\draw [->] (input) -- node[above] {$\mathbf{x}$} (sum);
	\draw [-] (sum) -- node[above, name=delta] {$\mathbf{\hat{w}}$} (pt1);
	\draw [->] (pt1) -- node[] {} (zM);
	\draw [->] (zM) -- node[above] {$\mathbf{u}$} (output);
	
	\node [point, below of=sum, node distance=1cm] (pt2) {};
	\node [point, below of=pt1, node distance=1cm] (pt3) {};
	\node [block, below of=delta, node distance=1.26cm] (IzR) {$I-\R$};
	\draw [-] (pt1) -- node {} (pt3); 
	\draw [-] (pt3) -- node {} (IzR);
	\draw [-] (IzR) -- node {} (pt2);
	\draw [->] (pt2) -- node[left] {$\mathbf{-\hat{x}}$} (sum);
	\end{tikzpicture}
	\caption{Controller implementation}
	\label{fig:blockdiag}
\end{figure}

Once we solve \eqref{eqn:synthesis} and obtain the optimal system responses $\left\{\R, \M\right\}$, we can implement a controller $\bK$ as per part 2 of Theorem \ref{thm:SLS}. Instead of directly inverting $\R$, we use the feedback structure shown in Fig. \ref{fig:blockdiag}, which is described by

\begin{equation}{\label{eqn:implementation}}
\begin{aligned}
        \mathbf{u}=\M \mathbf{\hat{w}},\ \ \  \mathbf{\hat{x}}=(\R-I)\mathbf{\hat{w}},\ \ \ 
        \mathbf{\hat{w}}=\mathbf{x}-\mathbf{\hat{x}}
\end{aligned}
\end{equation}

\noindent where $\mathbf{\hat{x}}$ can be interpreted as a nominal state trajectory, and $\mathbf{\hat{w}}=Z\mathbf{w}$ is a delayed reconstruction of the disturbance. This implementation is particularly useful because structural constraints (e.g. sparsity) imposed on the system responses $\left\{\R, \M\right\}$ translate directly to structure in the controller implementation. Thus, constraints on information sharing between controllers can be enforced as constraints on $\mathcal{X}$ and $\mathcal{U}$ in \eqref{eqn:synthesis}. Additionally, \eqref{eqn:implementation} allows for \textit{distributed implementation}. As mentioned above, each subsystem solves a local subproblem to obtain the local sub-matrices of $\left\{\R, \M\right\}$; these sub-matrices are then used to locally implement a sub-controller, independently of other subsystems.

\subsection{Model Predictive Control} \label{sec:mpc_math}
SLS-based MPC consists of solving the following problem at each timestep $t$:

\begin{equation} \label{eqn:mpc}
\begin{array}{rl}
\underset{\R, \M}{\text{min}} & f(\R, \M, x(t)) \\
\text{s.t.} &  \eqref{eqn:SLS_constraint}, \, \R \in \mathcal{X}, \M \in \mathcal{U} \\
 & \R x(t) \in \mathcal{P}_x \\
 & \M x(t) \in \mathcal{P}_u
\end{array}
\end{equation}

As with the core parametrization, $f$ is any convex objective function (e.g. $\mathcal{H}_2$), and $\mathcal{X}$ and $\mathcal{U}$ are convex sets that may specify communication-motivated sparsity constraints. $\mathcal{P}_x$ and $\mathcal{P}_u$ are convex sets that specify state and input constraints on predicted states and inputs, such as upper and lower bounds; these translate to affine constraints on $\left\{\R, \M\right\}$.

Equation \eqref{eqn:mpc} can be broken down into smaller subproblems to be solved in parallel using the alternating direction method of multipliers (ADMM) \cite{Boyd2010}. At each timestep, each subsystem accesses its own local subset of $x_t$, solves \eqref{eqn:mpc}, and outputs a local control signal $u_t$, which is calculated by multiplying the first block-matrix of $\M$ with $x_t$. Predicted values of $x$ and $u$ can be calculated by multiplying $x_t$ by the appropriate block-matrices of $\R$ and $\M$, respectively. This formulation enjoys the same scalability benefits as the core SLS method: runtime is independent of total system size. For a more thorough analysis of runtime, subproblem partitioning, etc., see \cite{AmoAlonso2019_mpc}.

	\section{SLS-BASED TECHNIQUES \& EXTENSIONS} \label{sec:sls_techniques}

We provide an overview of SLS-based works. The SLS parametrization provides a transparent approach to analyzing and synthesizing closed-loop controllers, and all methods described in this section exploit this fact. Some works focus on theoretical analyses, while others capitalize on the scalability provided by the SLS parametrization.

\subsection{Standard and Robust SLS}
The SLS parametrization was first introduced for state feedback in \cite{Wang2014_statefdbk}. For a comprehensive review of standard SLS, we refer the interested reader to \cite{Anderson2019} and reference therein; this work describes state and output feedback and relevant robust formulations. Though theory for the infinite-horizon case is well-developed, implementation is generally limited to finite-horizon approximations; this is partially remedied in \cite{Yu2020_h2}, which introduces an infinite-horizon implementation for the distributed state feedback $\mathcal{H}_2$ problem using SLS.

The main results on robust SLS are presented in \cite{Anderson2019}. Newer results on the robustness of the general SLS parametrization are found in \cite{Matni2020_robust}. Informally, robust SLS deals with cases in which the synthesized matrices $\left\{\R, \M\right\}$ do not describe the actual closed-loop system behavior, either by design or due to uncertainty. In these cases, robust SLS methods provide guarantees on the behavior of the closed-loop system. These guarantees are particularly useful for adaptation and learning, which are described later in this section.

In a setting with minimal uncertainty, robust SLS can be used when overly strict controller-motivated constraints on $\left\{\R, \M\right\}$ result in infeasibility during synthesis; this was the original goal of the robust SLS derivation. An alternative `two-step' approach is presented in \cite{Li2020_twostep}, whose results allow for separation of controller and closed-loop system constraints in the state-feedback setting.

\subsection{SLS for Nonlinearities \& Saturations}
SLS for nonlinear systems with time-varying dynamics is presented in \cite{Ho2020_nonlinear}. This work generalizes the notion of system responses to the nonlinear setting, in which they become nonlinear operators. No constraints are considered; instead, this work focuses on the relationship between achievable closed-loop maps and realizing state feedback controllers. 

Nonlinear SLS can be applied to saturating linear systems \cite{Yu2020_saturation} to provide distributed anti-windup controllers that accommodate state and input saturation constraints. An alternative approach to handling saturation is \cite{Chen2019_constraints}. This work uses robust optimization techniques instead of nonlinear analysis. In the $\mathcal{L}_1$ case with non-coupled constraints, the nonlinear method performs better; however, the linear method handles more general cases including coupling, using a distributed primal-dual optimization approach.

\subsection{Distributed Model Predictive Control}
An SLS-based distributed and localized MPC algorithm is developed in \cite{AmoAlonso2019_mpc}, and briefly described in the previous section. This the first closed-loop MPC scheme that allows for distributed sensing \textit{and} computation. Computation of this algorithm can be significantly accelerated via explicit solutions; so far, such solutions are available for the case of quadratic cost and saturation constraints \cite{AmoAlonso2020_explicitmpc}. Furthermore, this MPC algorithm readily extends to the robust setting -- this is the subject of current work. In addition to forming the basis for this MPC algorithm, the SLS parametrization can also be used to perform robustness analysis and guarantees on the general MPC problem \cite{Chen2019_robustmpc}.

\subsection{Adaptive Control \& Machine Learning}
A scalable adaptive SLS-based algorithm is presented in \cite{Ho2019_adaptive}. This work describes a framework for scalable robust adaptive controllers; at each timestep, measurements are collected to reduce uncertainty and boost performance while maintaining stability. This framework is applied in \cite{AmoAlonso2020_dropouts}, which introduces SLS controllers that are robust to package dropouts. A different scalable adaptive SLS-based algorithm deals with networks that switch between topological configurations according to a finite-state Markov chain \cite{Han2020_learning}.

SLS is especially beneficial for machine learning because it directly relates model uncertainty with stability and performance suboptimality \cite{Matni2019_rl}. The SLS parametrization is used to analyze the linear quadratic regulator (LQR) problem in the case where dynamics are unknown \cite{Dean2018_regret, Dean2019_complexity}. These works provide safe and robust learning algorithms with guarantees of sub-linear regret, and study the interplay between regret minimization and parameter estimation. Additionally, SLS forms the basis of a framework for constrained LQR with unknown dynamics, where system identification is performed through persistent excitation while guaranteeing the satisfaction of state and input constraints \cite{Dean2019_safety}. SLS is also used to provide complexity analysis and sharp end-to-end guarantees for the stability and performance of unknown sparse linear time invariant systems \cite{Fattahi2019_learning}. 

SLS-based analyses have been applied to the output feedback setting as well \cite{Dean2020_perception}. Motivated by vision-based control of autonomous vehicles, this work designs a safe and robust controller to solve the problem of controlling a known linear system for which partial state information is extracted from complex and nonlinear data. SLS also underpins the sample complexity analysis for Kalman filtering of an unknown system \cite{Tsiamis2020_kalman}. Additionally, SLS forms the basis for many ongoing works on data-driven control techniques.

\subsection{Additional works}
SLS is extended to the spatially invariant setting in \cite{Jensen2018_spatialinvar, Jensen2020_backstep}. In related work, \cite{Jensen2020_relativefull} explores realizable structured controllers via SLS and discusses the limitations of SLS in the relative feedback setting. Additionally, several works focus on optimizing computation for SLS. Explicit solutions to the general SLS problem are described in \cite{Anderson2018_structured}, and \cite{Tseng2020_dynamic} uses dynamic programming to solve for SLS synthesis problems 10 times faster than using a conventional solver. \cite{Tseng2020_deployment} describes deployment architecture for SLS, and is used in the construction of the SLSpy software framework \cite{Tseng2020_slspy}. In addition to this Python implementation, a MATLAB-based toolbox is also publicly available \cite{Li2019_SLSMatlab}. These toolboxes include implementations for standard and robust SLS, two-step SLS, nonlinear-saturation SLS, and distributed MPC.

	\section{CASE STUDY: POWER GRID} \label{sec:case_study}
We demonstrate the efficacy of a novel SLS-based controller on a power grid-like system with three key features:

\begin{enumerate}
	\item Periodic setpoint changes, induced by changing optimal power flows (OPFs) in response to changing loads
	\item Frequent, small\footnote{Relative to the size of the setpoint changes} disturbances
	\item Actuator saturation
\end{enumerate}

This system is inspired by common challenges of power systems control, though it is by no means a high-fidelity representation of the grid. The purpose of this case study is to show good performance in the presence of these common challenges in a simple and accessible system.

\subsection{System Setup}
\begin{figure}
    \begin{minipage}{4cm}
	\includegraphics[width=3.5cm]{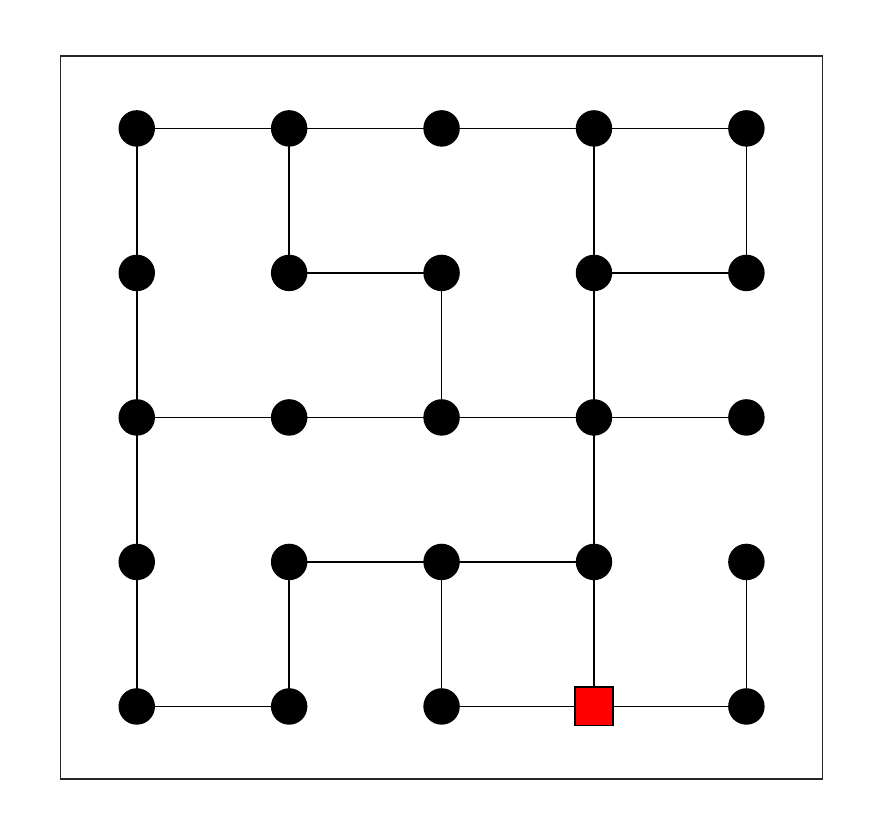}
	\end{minipage}
	\begin{minipage}{4cm}
	\caption{Topology of our example system for the case study. We will plot the time trajectories of states, disturbances, and input for the red square node.} \label{fig:topology}
	\end{minipage}
\end{figure}

We start with a randomly generated connected graph over a 5x5 mesh network, shown in Fig. \ref{fig:topology}. Graph edges represent connections between buses. Interactions between neighboring buses are governed through the undamped linearized swing equations \eqref{eqn:swing_eqn}, similar to the example used in \cite{Anderson2019}.

\begin{subequations} \label{eqn:swing_eqn}
\begin{equation} 
	x_i(t+1) = A_{ii} x_i(t) + \sum_{j \in \mathcal{N}(i)} A_{ij} x_j(t) + 
	\begin{bmatrix}
		0 \\ 1
	\end{bmatrix} (w_i(t) + u_i(t))
\end{equation}

\begin{equation} 
	A_{ii} = \begin{bmatrix}
		1 && \Delta t \\ -\frac{b_i}{m_i}\Delta t && 1
	\end{bmatrix}, \quad
	A_{ij} = \begin{bmatrix}
		0 && 0 \\ \frac{b_{ij}}{m_i}\Delta t && 0
	\end{bmatrix}
\end{equation}

\end{subequations}

The state of bus $i$ includes $x_i^{(1)}$, the phase angle relative to some setpoint, and $x_i^{(2)}$, the associated frequency, i.e. $x_i = [ x_i^{(1)} \quad x_i^{(2)} ]^{\top}$. $m_i$, $w_i$, and $u_i$ are the inertia, external disturbance, and control action of the controllable load of bus $i$. $b_{ij}$ represent the line susceptance between buses $i$ and $j$; $b_i = \sum_{j \in \mathcal{N}(i)}b_{ij}$. $b_{ij}$ and $m_i^{-1}$ are randomly generated and uniformly distributed between [0.5, 1], and [0, 10], respectively. Large values of $m_i^{-1}$ render the system unstable; the spectral radius of the $A$ matrix is 1.5.

\begin{figure}
	\centering
	\includegraphics[width=8.5cm]{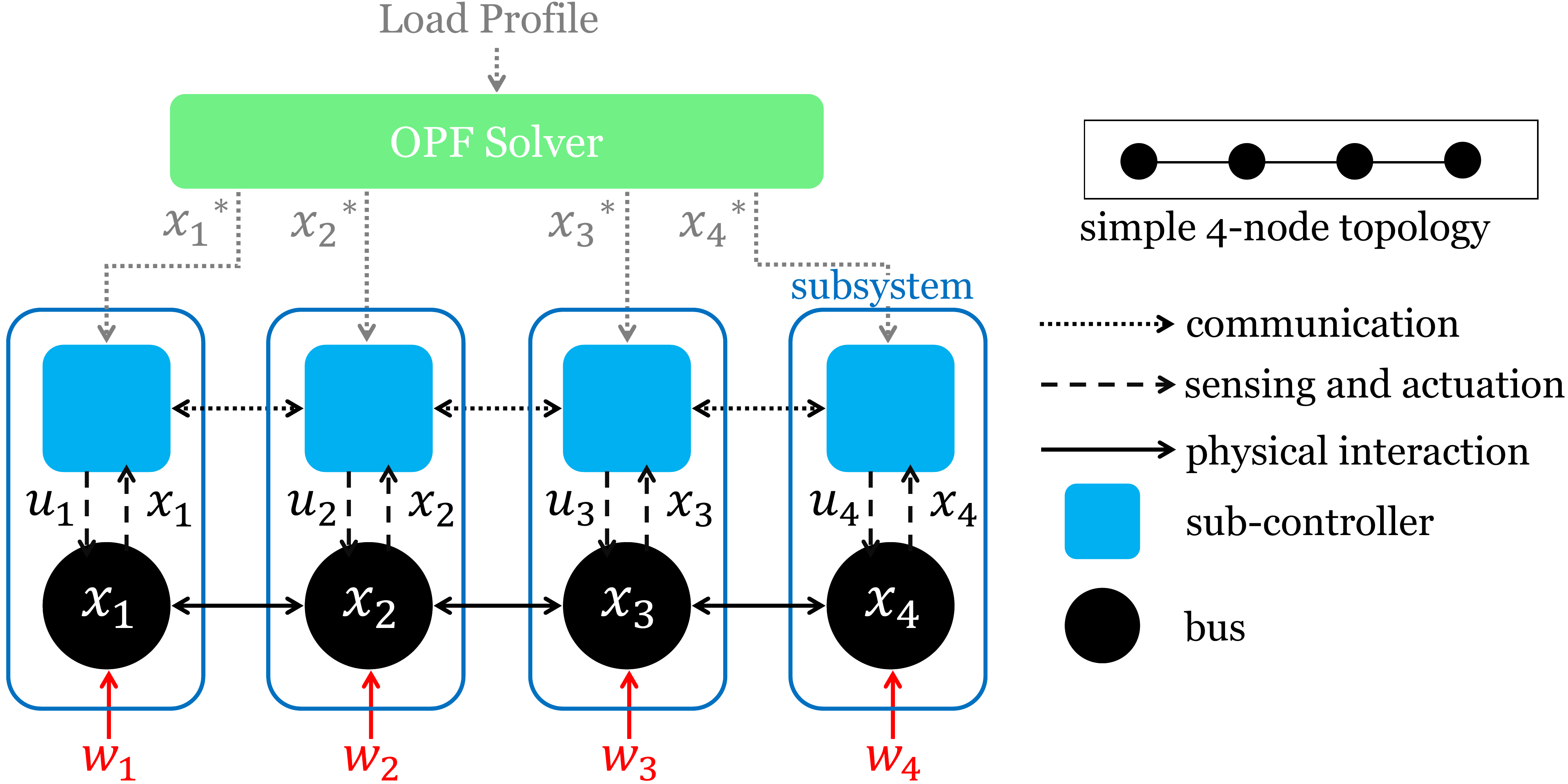}
	\caption{Architecture of example system. For ease of visualization, we depict a simple 4-node topology instead of the 25-node mesh we'll be using. Grey dotted lines indicate periodic communications; the OPF solver sends $x^*$ only on the timesteps when it receives a new load profile.} \label{fig:overall_arch}
\end{figure}

We periodically generate a new random load profile and solve a centralized DC-OPF problem to obtain the optimal setpoint $x^*$. 
We then send each sub-controller its individual optimal setpoint. Each subsystem reaches this setpoint in a distributed manner; sub-controllers are only allowed to communicate with their immediate neighbors and the neighbors of those neighbors (i.e. local region of size $d=2$). In addition to tracking setpoint changes, controllers must also reject random disturbances. The setup is shown in Fig. \ref{fig:overall_arch}.

We additionally enforce actuator saturation constraints of the form
$|u_i(t)| \leq u_{max}$. Actuator saturation is a ubiquitous constraint in control systems.

\subsection{Two-Layer Controller}

The combination of large setpoint changes, small disturbances, and relatively tight saturation bounds is challenging for an offline controller. Optimal performance requires the use of distributed SLS-based MPC. However, MPC incurs significant online computational cost since an optimal control problem must be solved at every timestep. We propose a two-layer controller to reduce this computation.

We decompose the main problem into two subproblems -- reacting to setpoint changes and rejecting disturbances -- and assign each subproblem to a layer. The top layer uses MPC to plan trajectories in response to large setpoint changes, while the offline bottom layer tracks this trajectory and rejects disturbances. Our controller offers unique scalability benefits compared to standard layered MPC controllers \cite{Scattolini2009}, and offers an efficiency boost over the SLS-MPC controller.

The top layer of the controller uses SLS-based MPC to generate saturation-compliant trajectories. As described in section \ref{sec:mpc_math}, this layer can be implemented in a distributed manner, with runtime independent of total system size. Every $T_{MPC}$ timesteps, the top layer solves \eqref{eqn:mpc} and generates a safe trajectory of states $x$ and inputs $u$ for the next $T_{MPC}$ timesteps -- this gives a $T_{MPC}$-fold reduction in computational cost compared to a single-layer MPC controller. To maximize performance, we time the top layer's trajectory generation to coincide with the periodic setpoint changes.

The trajectory generator alone is insufficient; since it only runs once every $T_{MPC}$ timesteps, disturbances can persist or amplify between runs, severely compromising performance. To deal with disturbances and preserve performance, we use an offline SLS controller in the bottom layer. This controller receives trajectory information from the top layer and outputs a control signal $u$ that tracks the desired trajectory.

\begin{figure}
	\begin{minipage}{4cm}
	\includegraphics[width=3.8cm]{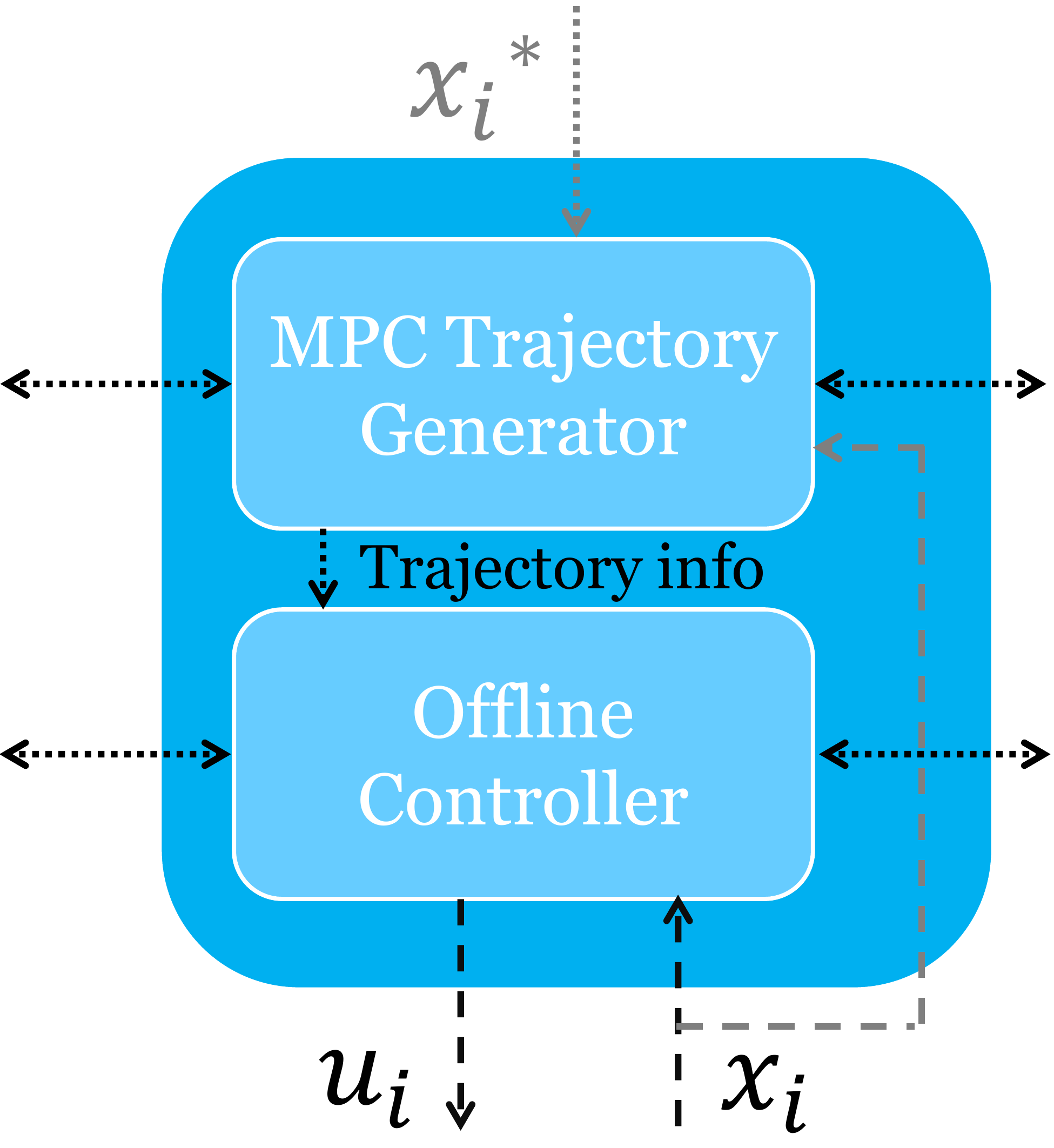}
	\end{minipage}
	\begin{minipage}{4cm}
	\caption{Layered sub-controller for the $i^{\text{th}}$ subsystem. Grey dotted line indicates periodic setpoint changes sent by the OPF solver. Grey dashed line indicates periodic sensing of local state $x_i$. Horizontal black dotted lines indicate communication with neighboring subcontrollers.} \label{fig:layered_ctrller}
	\end{minipage}
\end{figure}

The two-layer controller is shown in Fig. \ref{fig:layered_ctrller}. This layered controller is fully distributed; each subsystem synthesizes and implements both layers of its own controller, largely independently of other subsystems. Synthesis of the two layers is independent of one another, although some synthesis parameters may be shared. Additionally, all cross-layer communication is local, i.e. a top layer controller will never communicate with a bottom layer controller from a different subsystem.  Distributed implementation and synthesis allows our controller to enjoy a runtime that is independent of total system size, like its component SLS-based controllers. Thus, this layered MPC controller is uniquely scalable, and well suited for use in systems of arbitrary size.

We emphasize the contrast between our controller and the ideal controller -- centralized MPC. First, centralized MPC requires instantaneous communication between all subsystems and some centralized unit, while our controller is distributed and subsystems may only communicate within a local set of neighboring subsystems. Second, centralized MPC requires online computation once every timestep, while our controller requires online computation only once every $T_{MPC}$ timesteps. Third, centralized MPC's computational complexity scales quadratically with system size, which is impractical for large systems; our controller's complexity scales independently of system size. Despite this, our controller performs \textit{nearly identically} to this ideal centralized MPC. We summarize the above comparisons in Table \ref{table:cent_vs_loc}.

\begin{table}
	\caption{Centralized MPC vs. Local Layered MPC}
	\label{table:cent_vs_loc}
	\begin{center}
		\begin{tabular}{|l|l|l|}
			\hline
			& Centralized MPC & Local Layered MPC \\
			\hline \hline
			Communication & Global & Local \\
			\hline
			Online computation & Every timestep & Every $T_{MPC}$ timesteps \\
			\hline
			Complexity w.r.t. & \multirow{2}{*}{$O(N^2)$} & \multirow{2}{*}{$O(1)$} \\
			system size $N$ & & \\
			\hline
			Performance & Optimal & Within 1-3\% of optimal \\
			\hline
		\end{tabular}
	\end{center}
\end{table}
 
\subsection{Simulation Results}
We compare the performance of the layered controller with $T_{MPC}=20$ (`LocLayered') with the performance of the centralized MPC controller (`CenMPC'). For reference, we include the optimal linear controller with (`SatCenLin') and without (`UnsatCenLin') actuator saturation. The centralized controllers are free of communication constraints while the layered controller is limited to local communication.

\begin{figure}
	\centering
	\includegraphics[width=8.6cm]{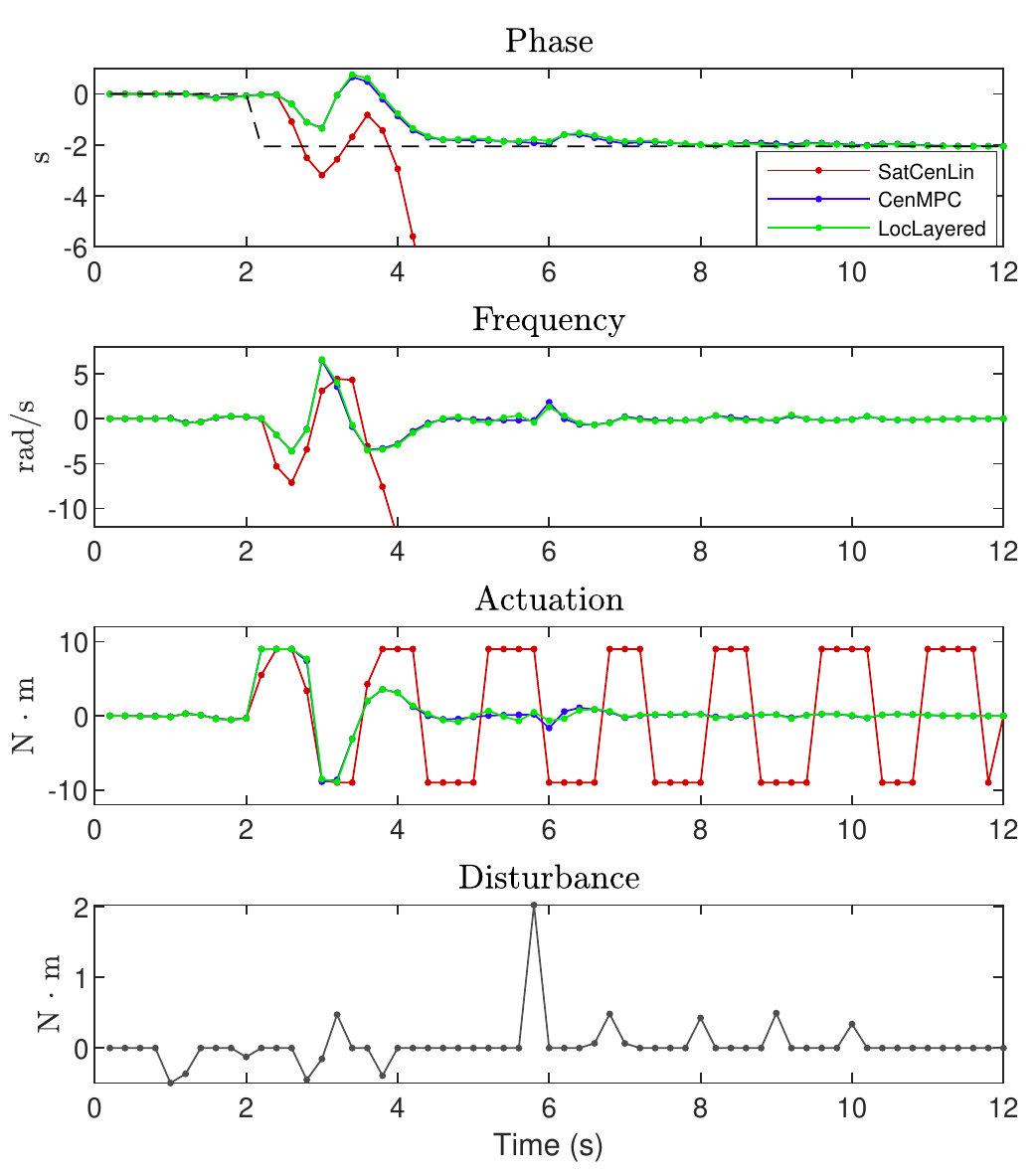}
	\caption{Performance of two centralized strategies (`SatCenLin', `CenMPC') and our layered controller (`LocLayered'). The linear saturating controller loses stability and gives extremely large oscillations in phase and frequency, which are omitted from the plot after around $t=4$; the associated actuation engages in oscillations as well, which are shown on the plot.}
	\label{fig:trajectory}
\end{figure}

The resulting LQR costs, normalized by the non-saturating optimal centralized cost, are shown in Table \ref{table:lqr_table_plotted}. We plot trajectories from the red square node from Fig. \ref{fig:topology} in Fig. \ref{fig:trajectory}, focusing on a window of time during which only one setpoint change occurs. The non-saturating controller is omitted.

\begin{table}
	\caption{LQR costs corresponding to Fig. \ref{fig:trajectory}}
	\label{table:lqr_table_plotted}
	\begin{center}
		\begin{tabular}{|l|l|l|l|}
			\hline
			\multirow{2}{*}{Controller} & Actuator & Global & \multirow{2}{*}{LQR Cost} \\
			& Saturation & Communication & \\
			\hline \hline
			UnsatCenLin & No & Yes & 1.00 \\
			\hline
			SatCenLin & Yes & Yes & 3.97e13 \\
			\hline			
			CenMPC & Yes & Yes & 1.90 \\
			\hline
			LocLayered & Yes & No & 1.93 \\
			\hline
		\end{tabular}
	\end{center}
\end{table}

For this setpoint change, saturation-induced windup effects result in oscillations of increasing size from the saturated linear controller, causing it to lose stability and incur astronomical costs. Windup effects are mitigated by both online controllers, which perform similarly despite drastically different computational requirements. As desired, the proposed layered controller achieves near-optimal performance. 

To check general behavior, we re-run the simulation 30 times with different randomly generated grids, plant parameters, disturbances, and load profiles. The resulting LQR costs, normalized by the non-saturating optimal cost in each run, are shown in Table \ref{table:lqr_table_average}. We show the mean cost over all 30 trials and the mean cost over the 26 trials in which the saturated linear controller maintained stability. Observations from the initial example hold; the layered controller consistently achieves near-optimal performance. The two online methods again demonstrate enormous improvement over the saturated linear controller, which often loses stability. Performance differences predominately arise from reactions to setpoint changes; when the saturated linear controller manages to maintain stability after a setpoint change, performance is similar across all methods. Lastly, we note that both the online methods' costs are within 17\% of the centralized optimal cost \textit{without} actuator saturation.

\begin{table}
	\caption{LQR costs averaged over 30 trials}
	\label{table:lqr_table_average}
	\begin{center}
		\begin{tabular}{|l|l|l|}
			\hline
			\multirow{2}{*}{Controller} & \multicolumn{2}{|c|}{LQR cost} \\
			\cline{2-3}
			 & Total \qquad \qquad & SatCenLin stable \\
			\hline \hline
			UnsatCenLin & 1.00 & 1.00 \\
			\hline			
			SatCenLin  & 1.32e7 & 1.13 \\
			\hline
			CenMPC & 1.16 & 1.08 \\
			\hline
			LocLayered & 1.17 & 1.09 \\
			\hline			
		\end{tabular}
	\end{center}
\end{table}

	\section{CONCLUSIONS} \label{sec:conclusions}

We reviewed SLS and derivative works, and demonstrated an effective combination of SLS-based methods in a novel layered controller. This controller performed similarly to centralized MPC and is superior in scalability, communication requirements, and computational cost. 

SLS-based methods are effective standalone controllers. However, as the systems we seek to control become increasingly complex and uncertain, standalone controllers are insufficient; techniques must be combined (i.e. via layering) to create a controller that is at once safe, scalable, efficient, and high-performing. SLS-based methods are excellent candidates for use in such layered controllers, as they readily interface with one another and with learning-based methods.
	
	\bibliography{sls, other}
	\bibliographystyle{IEEEtran}
\end{document}